\documentclass[twoside]{article}
\usepackage[latin1]{inputenc}
\usepackage{t1enc}
\usepackage{a4}
\usepackage{tabularx}
\usepackage{epsf}

\def\bref{\vspace{4pt}\noindent\hangindent=10mm}
\def\farcs{\hbox{$.\!\!^{\prime\prime}$}}

\def\degr{\hbox{$^\circ$}}
\def\sun{\hbox{$\odot$}}
\def\kms{\hbox{km\,s$^{-1}$}}
\def\fm{\hbox{$.\!\!^{\rm m}$}}
\def\Msun{\hbox{M$_{\sun}$}}
\def\NH{\hbox{[N\,{\sc ii}]$\lambda$6583\AA/H$_{\alpha}$}}

\def\arcmin{\hbox{$^\prime$}}
\def\arcsec{\hbox{$^{\prime\prime}$}}

\begin{document}

\setcounter{figure}{0}
\setcounter{section}{0}
\setcounter{equation}{0}

\begin{center}
{\Large\bf
LBV Nebulae: The Mass Lost from\\[0.2cm]
the Most Massive Stars}\\[0.7cm]
 
Kerstin Weis \\[0.17cm]
Institut f\"ur Theoretische Astrophysik \\
Universit\"at Heidelberg\\
Tiergartenstr. 15\\
69121 Heidelberg\\
kweis@ita.uni-heidelberg.de\\
http://www.ita.uni-heidelberg.de/{$\sim$}kweis\\
\end{center}

\vspace{0.5cm}

\begin{abstract}
\noindent{\it

The most massive stars, with initial masses above $\sim$
50\,\Msun, encounter a phase of  extreme mass loss---sometimes 
accompanied by so-called giant eruptions---in which the stars'
evolution is reversed from a redward to a blueward motion in the HRD.
In this phase the stars are known as {\it Luminous Blue Variables} (LBVs).
Neither the reason for the onset of the strong mass loss nor the 
cause for the giant eruptions is really understood, nor is their 
implications for the evolution of these most massive stars.
I will present a study of the LBV nebulae which are formed in this 
phase as a consequence of the strong mass loss and draw conclusions 
from the morphology and kinematics of these nebulae on possible 
eruption mechanisms and stellar parameters of the LBV stars.
The analysis contains a large collection of LBV nebulae
which form an evolutionary sequence of LBV nebulae.
A special concern will be the frequently observed bipolar nature of the 
LBV nebulae which seems to be a general feature and presents strong constraints
on further models of the LBV phase and especially on the formation mechanism 
of the nebulae.  
}
\end{abstract}

\section{Introduction}

Due to their brightness, massive stars can be investigated as individual 
objects even in neighboring galaxies. They have initial masses between 
$\sim$ 15\,\Msun\ and 120\,\Msun, bolometric luminosities 
of 10$^{5}$ to 10$^{6}$ L$_{\sun}$ and they live about $2-10$ 10$^6$ years
(eg., Schaller et al.\ 1992, Maeder \& Conti 1994, Schaerer et
al.\ 1996a, 1996b).
Through stellar winds these stars loose a sizeable fraction of their initial
mass during their life (already more than 50\% during the main-sequence phase)
corresponding to an average mass loss rate of 
10$^{-5...-6}$ \Msun yr$^{-1}$ (eg., de Jager et al.\ 1988, Kudritzki 
et al.\ 1996, Kudritzki 1999) which compares to the loss of low mass stars 
like the Sun of $\sim$ 10$^{-14}$ \Msun yr$^{-1}$.   
Only in recent years it has become obvious that this mass loss 
strongly influences the stellar evolution (eg., Chiosi \& Maeder 1986)
and both, directly and due to the 
change in evolution, has strong implications on the evolution of the 
interstellar and circumstellar environment.
Massive stars therefore not only are major contributors of
high energy photons which excite for example H\,{\sc ii} regions,
they also---through their winds---supply a large amount of kinetic 
energy into the surrounding medium. Their largest input, however, is still 
given as these stars explode as supernovae at the end of their evolution.   

Stars with masses  above $M \sim 50$\,M$_{\sun}$ and luminosities
of $L\sim 10^{6}$\,L$_{\sun}$ top the {\it Hertzsprung-Russell
Diagram\/} (HRD) and represent the most massive stars known. After
spending a `normal' life as O stars on the main sequence, they
evolve towards cooler temperatures and enter a phase with---even for their 
standards---{\it extremely} high mass loss (up to 
10$^{-3...-4}$\,M$_{\sun}$yr$^{-1}$) and become
{\it Luminous Blue Variables\/} (LBVs; Langer et al.\ 1994, Humphreys 
\& Davidson 1994, Maeder \& Meynet 2000). In the
HRD, the LBV phase starts when the stars reach the empirical {\it
Humphreys-Davidson limit\/} (Humphreys \& Davidson 1979, Humphreys \& 
Davidson 1994). It
seems as if the most massive stars do not evolve beyond this
limit towards lower temperatures. They rather turn around,
evolve back and become blue supergiants located in the HRD in the
vicinity of the Humphreys-Davidson limit. Moreover, this is the
location in the HRD where those stars undergo extremely strong mass
loss through winds and occasional giant eruptions (eg., Humphreys 
1999), and thus peel off parts of the stellar envelope. The LBV stars
therefore often form small circumstellar nebulae, the {\it LBV
nebulae\/} (eg., Nota et al.\ 1995). For a compilation of 
LBV nebulae known today, see Table \ref{table:lbvn}. The formation 
of the nebulae as
well as the consequences of the mass loss on the evolution of the
most massive stars, especially in the LBV phase, is still far from
being understood. In particular it is neither clear what causes
the strong mass loss to set in, and thus to reverse the direction
of the star's evolution in the HRD, nor what leads to the giant
eruptions. Their closeness to the classical Eddington 
limit, however, makes a loss of radial balance not very surprising.

First hydrodynamic calculations modeled the formation 
of LBV nebulae through the interaction of slow and fast winds in the 
stars' evolution. The model by Garc{\'\i}a-Segura et al.\ (1996), for 
instance, was calculated for a 60\,\Msun\ star, and shows that the 
slow wind in the beginning of the LBV phase 
is swept up by the following faster wind and compressed into a 
thin shell, which manifests itself as an LBV nebula. 
This wind-wind interaction model predicts an expansion velocity 
of about $\sim$ 240\,\kms\ for the nebula and that, due to the mixing up and
peeling off of material from the CNO cycle, the nitrogen abundance is enhanced
by a factor of about $\sim 13$. While this model may be able to explain
the formation of LBV nebulae formed by stellar wind, LBV nebulae
which were created in a giant eruption these models are 
not able to account for. In giant eruptions the LBVs increase their 
brightness by typically $2-5$ magnitudes within only a few years
(see for example Fig.\,1(b) in Humphreys et al.\ 1999 for P Cygni). 
The total energy output during such an event is $\sim$ 10$^{50}$
ergs and therefore nearly comparable to a supernova---actually 
there were two giant eruption LBVs which were originally mistaken for
supernova explosions (Humphreys 1999), namely SN\,1961V (in NGC 1058) 
and SN\,1954J (in NGC 2403).
Simple wind-wind interaction models are not able to get such outbursts and
the formation of the LBV nebula on such short timescales. 
The outburst phases are 
not longer than a few years, while the formation of a
nebula due to interaction of winds takes at least several 10$^3$ years, 
according to the model of Garc{\'\i}a-Segura et al.\ (1996).

\begin{table}
\caption[]{Parameters of known LBV nebulae}\label{table:lbvn}
\begin{center}
\begin{tabular}{cccccc}
\hline
\hline
LBV & host galaxy & size & $v_{\rm exp}$ & references \\
& & [pc] & [\kms] & \\
\hline
\hline
$\eta$ Carinae & Milkyway & 0.2/0.67 & 600/$10-2000$ & 1, 2 \\
HR Carinae & Milkyway & 1.3\,$\times$\,0.65 & $75-150$ & 3, 4\\
P Cygni & Milkyway & 0.2/0.8 & $110-140$/185 & 5 \\
AG Carinae  & Milkyway & 0.87\,$\times$\,1.16 & 70 & 6 \\
WRA 751 & Milkyway & 0.5 & 26 & 7 \\
He 3-519 & Milkyway & 2.1 & 61 & 8 \\
HD 168625 & Milkyway & 0.13\,$\times$\,0.17& 40 & 9 \\
Pistol Star & Milkyway & 0.8\,$\times$\,1.2 & 60 & 10 \\
R127 & LMC & 1.3 & 32 & 11 \\
R143 & LMC & 1.2 & 24 (?) & 11 \\
S61 & LMC & 0.82 & 27 & 11 \\
S119 & LMC (?) & 1.8 & 26 & 12 \\
\hline
Sher 25 & Milkyway & 1 & 70 & 13, 14 \\
Sk$-69\degr$ 279 & LMC & 4.5 & 14 & 15, 16\\
\hline
\end {tabular}
\begin{tabular}{l}
(1) Davidson \& Humphreys (1997), (2) this work, (3) Weis et al.\\ (1997a), 
(4) Nota et al.\ (1997), (5) Meaburn et al.\ (1996a), (6) Nota \\ et  al. 
(1992), (7) Weis (2000a), (8) Smith et al.\ (1994), (9) Nota et \\ al. 
(1996), (10) Figer et al.\ (1999), (11) Weis (2000b), (12) Weis et \\ al. 
(2000a) , (13) Brandner (1997a), (14)
Brandner (1997b), (15) \\ Weis et al.\ (1997b), (16) Weis \& Duschl (2000).\\
\hline
\hline
\end{tabular}
\end{center}
\end{table}

As yet, only a few LBVs are known in our Galaxy (8 objects plus
candidates) and some more in other galaxies, like the Large Magellanic 
Cloud (LMC), M31 and M33 (Humphreys \& Davidson 1994). Altogether 
around 40 LBV stars are currently known, many of which show circumstellar 
nebulae. To better understand the LBV phase and especially the formation of 
the LBV nebulae, and the onset of the enhanced mass loss, in order to 
address the question what causes  the evolution of the most massive stars
to reverse, I have analyzed the nebulae of most of the LBVs known. 
What increases the mass loss
in the LBV phase? What drives the giant eruptions, what mechanism forms 
instabilities that might lead to the eruptions? And, how do the nebulae form? 
Only if we can answers these questions we well be able to better understand
the structure and evolution of the most massive stars.

\section{The Nebula around the LBV $\eta$ Carinae}

\subsection{Historical Background}

\begin{figure}[t]
\epsfxsize=0.95\textwidth
\centerline{\epsffile{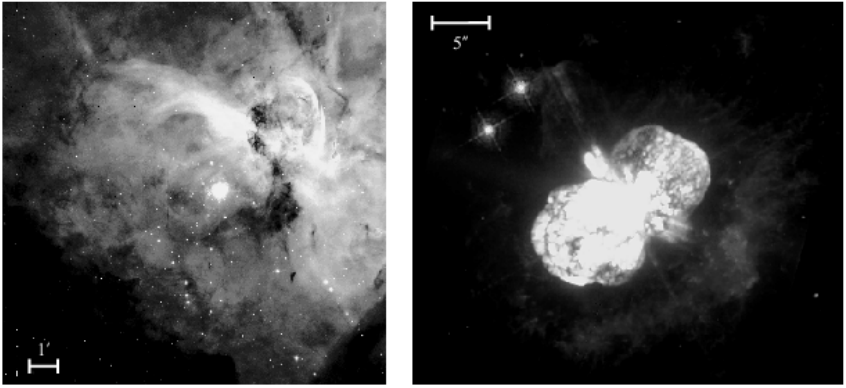}}
\caption{{\it Left:} This figure shows an 
H$_{\alpha}$ image, taken with the CTIO 0.9\,m telescope 
of the central region in NGC 3372, called the Keyhole nebula. The 
keyhole shaped nebula is visible as well as a bright, extended source 
to the east which is the LBV $\eta$ Carinae and its nebula. The field of 
view is about 13\arcmin\,$\times$\,13\arcmin, east is left and north 
to the top. {\it Right:} An HST image reveals clearly the
bipolar nature of the Homunculus nebula around $\eta$ Carinae.
Again east is left and north to the top.} 
\label{fig:etahst}
\end{figure}

$\eta$ Carinae, perhaps the best-known of all LBVs, is located in NGC 3372,
one of the largest H\,{\sc ii} regions in our Galaxy. It is often referred 
to as the {\it Carina Nebula}. 
$\eta$ Carinae belongs to the Trumpler 16 cluster
(Massey \& Johnson 1993, Walborn 1995) which is close to the 
central region of NGC 3372 which, in turn is known as the  
{\it Keyhole nebula} (see Fig.\,\ref{fig:etahst}), at a 
distance of about 2.3\,kpc from us (Walborn 1995).
$\eta$ Carinae has a bolometric luminosity of 
$L = 10^{6.7}$\,L$_{\sun}$ which puts the star among the most luminous 
stellar objects
and---with an estimated mass of about 120\,M$_{\sun}$---also among the 
most massive stars in our Galaxy (Humphreys \& Davidson 1994, Davidson \& 
Humphreys 1997). Even if recent indications of a binary nature
of $\eta$ Car (see Damineli 1996, Damineli et al.\ 1997, 
Stahl \& Damineli 1998) prove to be true, at least one component
must be within the range of LBVs, i.e., larger than $\sim$50\,M$_{\sun}$.
The binary hypothesis is highly under debate still. Since it is not 
a crucial constraint for the work presented here this 
issue will not be addressed further. For detailed discussions, we refer the
reader to the above mentioned references.

Historical  observations of $\eta$ Carinae 
showed that the star is variable and changed between 4$^{\rm m}$ and 
2$^{\rm m}$ 
in the 17th and 18th century. The star's most peculiar behavior, 
however, was its outburst around  1843 when it
increased its visual magnitude to $-$1$^{\rm m}$ for a short time 
($\sim$10 years) before its luminosity declined by 7 magnitudes in only a 
few years---$\eta$ Carinae had undergone a giant eruption. For lightcurves, 
see, for example, Gratton (1963), Viotti (1995), and Humphreys (1999).

During this eruption a circumstellar LBV nebula had formed which
was not detected until 1938 when van den Bos noted a {\it nebula halo} 
when observing components of $\eta$ Carinae which he thought were parts 
of a multiple system. Only around 1950 it became clear that these 
components were brighter knots of a fainter nebula. Nearly 
simultaneously but independently Gaviola (1946, 1950) and Thackeray (1949,
1950) discovered and photographed this nebula for the first time. Due to the 
odd man-like shape of the nebula Gaviola named it
the {\it Homunculus\/}. Gaviola also noted that there may be a much fainter 
and larger hook-shaped nebula surrounding the Homunculus.
Indeed there was an outer nebula. Deeper images by Walborn (1976) 
showed such a larger nebula which consists of structures 
as, for instance, the {\it S ridge\/}, the {\it E condensations\/}, or the
{\it W arc \/} (for a complete description see the sketch by Walborn 1976).
It was not until 1995 that the first high-resolution images were published: a
ground based image (Duschl et al.\ 1995), which was already taken 
in 1985 and re-constructed by a {\it shift-and-add-like} technique, 
and the first (COSTAR corrected) HST images taken in 1993 
(eg., Morse et al.\ 1998). The images 
reveal that the morphology of the Homunculus nebula 
is highly bipolar. Two lobes are separated by an equatorial disk (Duschl et
al. 1995) which is defined by structures called streamers. The HST image 
in Fig.\,\ref{fig:etahst}  shows very well the bipolar 
nebula and the equatorial
disk. A careful inspection of the images reveals that the Homunculus shaped
nebula, reported earlier, manifests only the brightest 
structures within the bipolar nebula. An even longer exposure 
(200\,s) with the HST (see Fig.\,\ref{fig:hstlong}) in the 
F656N (H$_{\alpha}$) filter detects a large amount of 
filaments and knots in addition to the structures already known
outside of the Homunculus (as reported by Walborn 1976). 

First measurements of the expansion of the Homunculus nebula were published 
by Ringulet (1958) and Gratton (1963). They reach to values of 
500\,\kms\ for the expansion (corrected to today's distance determination
of $\eta$ Car). A very detailed analysis of proper motion measurements, again
for some of the outer filaments, has been made in a 
series of studies by Walborn (1976),
Walborn et al.\ (1978) and Walborn \& Blanco (1988), in which they find 
tangential velocities between 280\,\kms\ and 1360\,\kms. Recent proper 
motion measurement using the high-resolution HST images support these 
results: Currie et al.\ (1996) and Currie \& Dowling (1999) obtain values 
between 10 and 1000\,\kms\ as do Smith \& Gehrz (1998) through a comparison of 
old ground based images with a resolution-degraded HST image, 
enlarging the time base. 
Thackeray (1961) was among the first to derive radial velocities from 
spectra. Comparing forbidden and permitted lines he
concluded that the Homunculus expands with 630\,\kms. Later studies using a
long-slit mode show even higher values for structures outside 
the Homunculus. Meaburn et al.\ (1987, 1993a, 1993b, 1996b) report on radial
velocities from at least 250\,\kms\ to $-1200$\,\kms, with an additional
uncertain feature at $-1450$\,\kms. Fabry-Perot 
measurements (Walborn 1991) agree with these values.

\subsection{Morphology}

\begin{figure}[th]
\epsfxsize=0.95\textwidth
\centerline{\epsffile{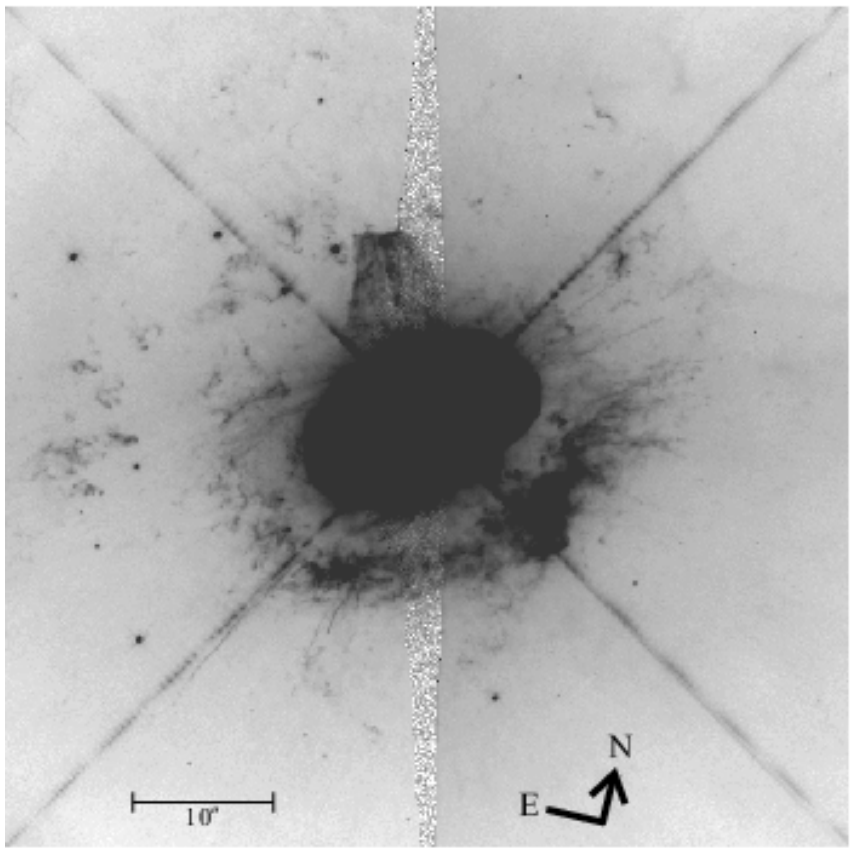}}
\caption{This figure shows an HST image taken with the F656N filter (H$_{\alpha}$)
of the nebula around the LBV $\eta$ Carinae. The field of view is about
60\arcsec\,$\times$\,60\arcsec.} 
\label{fig:hstlong}
\end{figure}

In order to analyze the morphology of the nebula around $\eta$ Carinae 
we used deep high-reso\-lu\-tion images from the HST archive. 
Figure\,\ref{fig:hstlong} shows a  F656N filter image. 
Note that this image is showing not only H$_{\alpha}$ emission but in addition 
red- and blueshifted emission from the [N\,{\sc ii}] lines at 6548\,\AA\ and
6583\,\AA. Due to the large Doppler shift (above 1000\,\kms) of at least some
of the filaments, parts of the nebula's emission in the F656N filter is 
ambiguous. The brightness levels displayed in this image are optimized 
to especially reveal faint structures in the outskirts of the nebula.
The bipolar Homunculus is unresolved as the oval shaped saturated 
central part. While the Homunculus measures about 17\arcsec\ 
(about 0.2\,pc) across, fainter structures are detected up to  
distances of 30\arcsec\ (0.33\,pc) from the star. All knots outside the 
Homunculus taken together form the
{\it outer ejecta\/}. Knots and condensations classified earlier by Walborn 
(1976) are part of these  outer ejecta.  

The knots and filaments in the outer ejecta consist of very different
morphological shapes and sizes. They can be described as bullets, long
filaments, some even look like arches---but one has to keep in mind 
that all structures are only seen in projection. The sizes of 
coherent structures
vary from as large as 7\farcs5 (0.08\,pc, the {\it N Condensation}\/) down 
to fractions of arcseconds (several 10$^{-3}$\,pc)---the 
limit of what HST can resolve. 
Most likely the outer ejecta contains numerous even smaller and/or fainter
structures below the detection limit of this image.
One of the most striking features are very straight, long and collimated 
objects which in the following we will call {\it strings}. These structures
will be discussed in Section\,\ref{sect:strings}. 

An area in the outer ejecta particularly rich in structures like bullets and 
filaments is the {\it S ridge} (with the brightest part called the {\it S
Condensation}, Walborn 1976). 
The deep HST image (Fig.\,\ref{fig:hstlong}) shows in 
great detail the countless  
knots present in a nearly circular ring around $\eta$ Carinae with 
long filaments pointing roughly radially outwards in this ridge.
An analysis of the X-ray emission from the nebula around $\eta$ Carinae
(Weis et al.\ 2000b) supports that most likely high-velocity knots colliding 
with their environment  are present in the {\it S ridge} 
and give rise at this position to the highest 
X-ray emission from the nebula.

\subsection{Kinematics and 3D Structure}

For a detailed study of the kinematics of the nebula---which 
will help to better model and understand its 3 dimensional 
structure---high-resolution echelle long-slit observations were made, 
using the Cerro 
Tololo Inter-American Observatory's 4\,m telescope. The instrumental 
FWHM of the spectra at the H$_{\alpha}$ line was about 14\,\kms. 
With spacings of the offsets chosen to match  the seeing (1\farcs5)
a complete 2 dimensional mapping was secured, taking 31 spectra 
and covering an area of about 1\arcmin\,$\times$\,1\arcmin\ 
(0.67\,pc\,$\times$\,0.67\,pc). 
A position angle (PA) of 132\degr\ was chosen, putting the slits 
parallel to the major axis of the Homunculus. Due to strong stray-light from
the central star the central 6 slit positions could not be used.

\begin{figure}
\epsfxsize=0.95\textwidth
\centerline{\epsffile{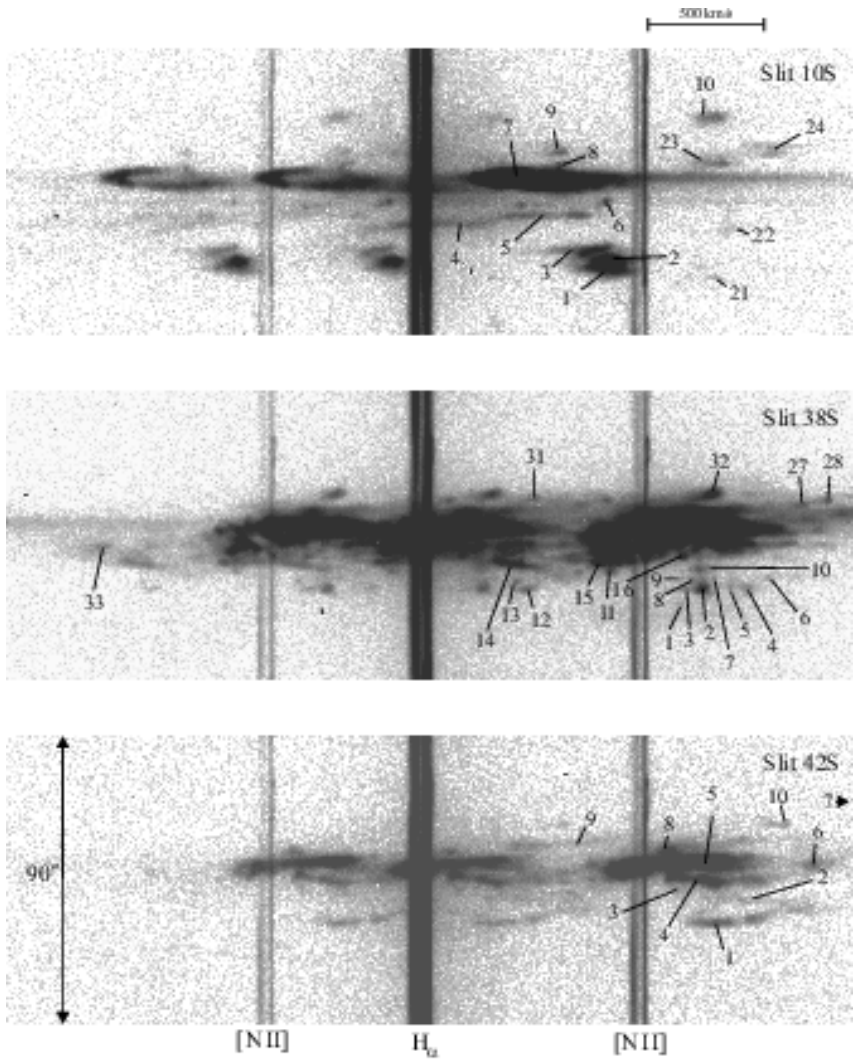}}
\caption{The echellograms displayed here are 90\arcsec\ high in spatial
  direction and cover a spectral region of 75\,\AA\ centered on
  rest-wavelength H$_{\alpha}$.
In many cases redshifted  as well as blueshifted knots appear simultaneously,
producing the `messy spectra'. Since all knots are present in all 3 lines
included in the spectra an unambiguous identification was nevertheless 
possible for most of them.} 
\label{fig:slits}
\end{figure}

Figure\,\ref{fig:slits} depicts 3 echellograms taken at different positions.
The spatial axis in each echellogram is 90\arcsec\ high, the 
spectral axis covers 75\,\AA\ centered on H$_{\alpha}$ at its rest wavelength. 
Beside the splited line emission (split by about 40\,\kms) across the 
entire spectrum, which results from the background H\,{\sc ii} region, 
a large variety of blue- and redshifted knots from the nebula 
around $\eta$ Carinae is visible in the spectra (Fig.\,\ref{fig:slits}).

The maximum and minimum velocities of each knot identified
in the spectra were measured, together with the velocities of the brightest
intensity maxima of a knot. The knots as identified from the spectra
were compared with and re-identified in the HST images. 
More than 200 individual
knots in the spectra were found and, as seen in Fig.\,\ref{fig:slits},
nearly all of them show  substructures. A knot is defined as a single
entity if it shows a coherent appearance in the spectra. For example 
knot\,1 in Slit 10S is a single feature as is knot\,7 in Slit 10S. 
Substructure means that these 
structures are coherent but show different intensities or a change in the
velocity profile (e.g. knot\,7 in Slit 10S). 
Taking all spectra together we obtain the following overall properties: 
The fastest blueshifted structure moves with a maximum radial
velocity of $-1200$\,\kms, the fastest redshifted features reach up to
$+2000$\,\kms. The majority of the knots shows radial velocities reaching 
between $-600$\,\kms\ and $+600$\,\kms\ (would these structures have a
projection angle of 45\degr\  the  unprojected
velocities were already between  $-850$\,\kms\ and $+850$\,\kms).    
The largest velocity spread of one single coherent structure is as high as
1250\,\kms.    

The spectra also allow us to obtain the line ratio \NH\ leaving a hint for
the strength of nitrogen within the nebula. As stated above an enhanced
nitrogen ratio is expected due to the overabundance of nitrogen in the nebula
which results from the CNO cycled material peeled off from the star. For the 
structures measured in the outer ejecta this \NH\ ratio shows values 
between $1-5$ with a peak around 3.

\begin{figure}
\epsfxsize=0.8\textwidth
\centerline{\epsffile{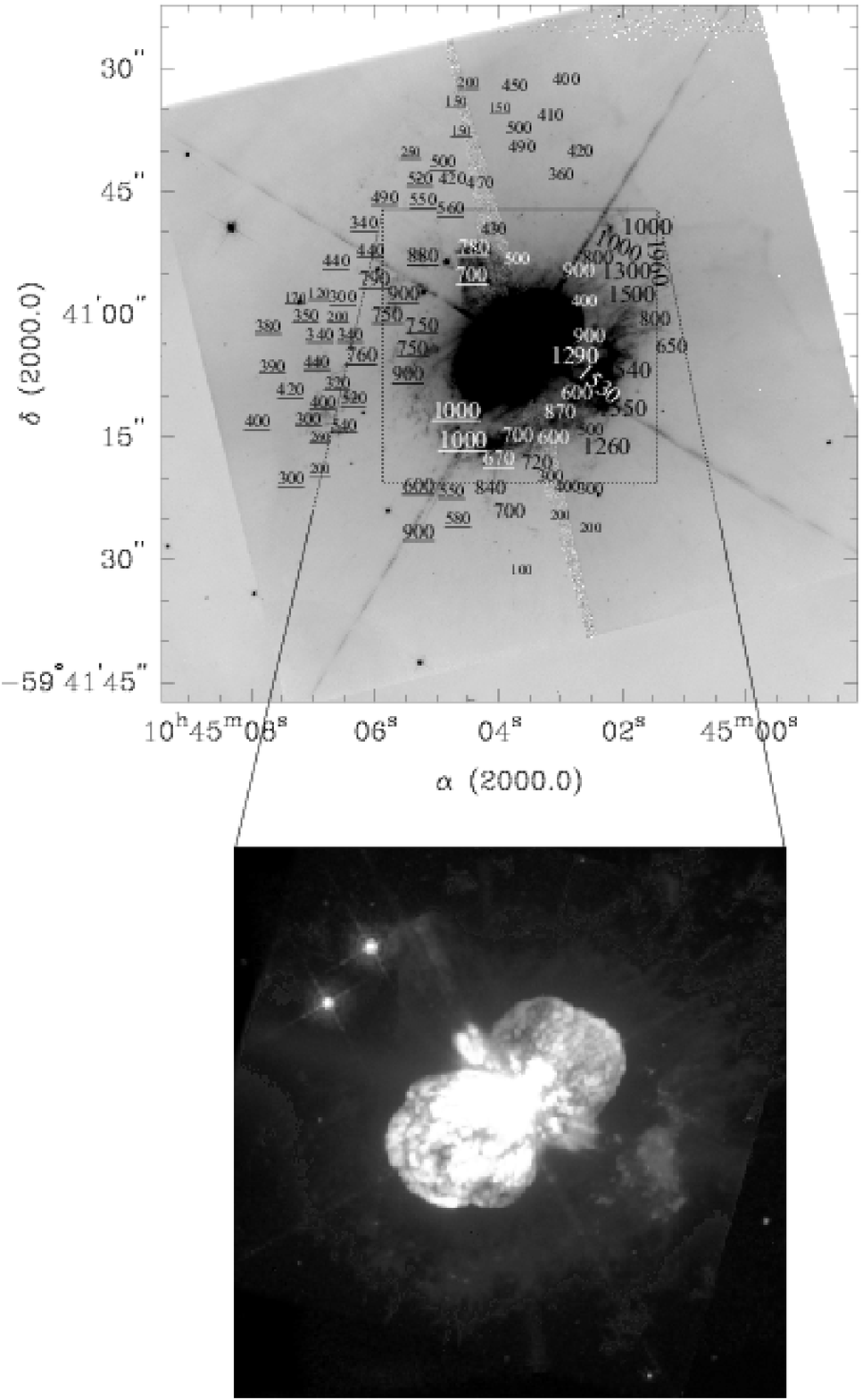}}
\caption{In the upper panel the same image as in Fig.\,\ref{fig:hstlong} 
is shown. 
Onto the image, scaled by font sizes, the radial velocities of individual
knots (with the highest velocity for a certain area) are overplotted.
Underlined velocities represent negative (blushifted) structures,
not underlined are redshifted, positive values. A clear trend is visible:
blueshifted structures appear preferentially in the south-east while redshifted
are concentrated to the north-west. The lower panel shows the central 
bipolar Homunculus for comparison. In both images north is up and east to the
left.} 
\label{fig:etabipolar}
\end{figure}
 
Beside the kinematic information for each individual structure  the
measurements show an interesting global result. 
An overlay of respective
knots onto an HST image as given in Fig.\,\ref{fig:etabipolar}
clearly shows the trend that blueshifted structures appear 
in the south-east while redshifted are concentrated to the north-west.
The knots in the outer ejecta---which seem to be randomly distributed 
around $\eta$ Carinae---seem to follow an ordered 
kinematic pattern. They are moving in a bipolar mode
with respect to the central star. Knots in the south-east are approaching us, while knots north-west of 
$\eta$ Carinae move away from the observer. Going even further and comparing
the bipolar structure of the outer ejecta with the inner central bipolar
Homunculus we can see that both parts of the nebula follow the same symmetry.
The  south-eastern  lobe of the Homunculus is tilted towards the observer 
(the net expansion here is therefore blueshifted---approaching) while the 
north-western lobe is tilted to the back (redshifted---receding).
This bipolarity is repeated in the filamentary 
outer ejecta with the same symmetry and with about the same symmetry 
axis, i.e., both the Homunculus and the outer ejecta exhibit 
a very similar bipolar 
structure of the nebula around $\eta$ Carinae. 

\subsection{Strings---a new Phenomena in LBV Nebulae? }\label{sect:strings}

\begin{figure}
\epsfxsize=0.47\textwidth
\centerline{\epsffile{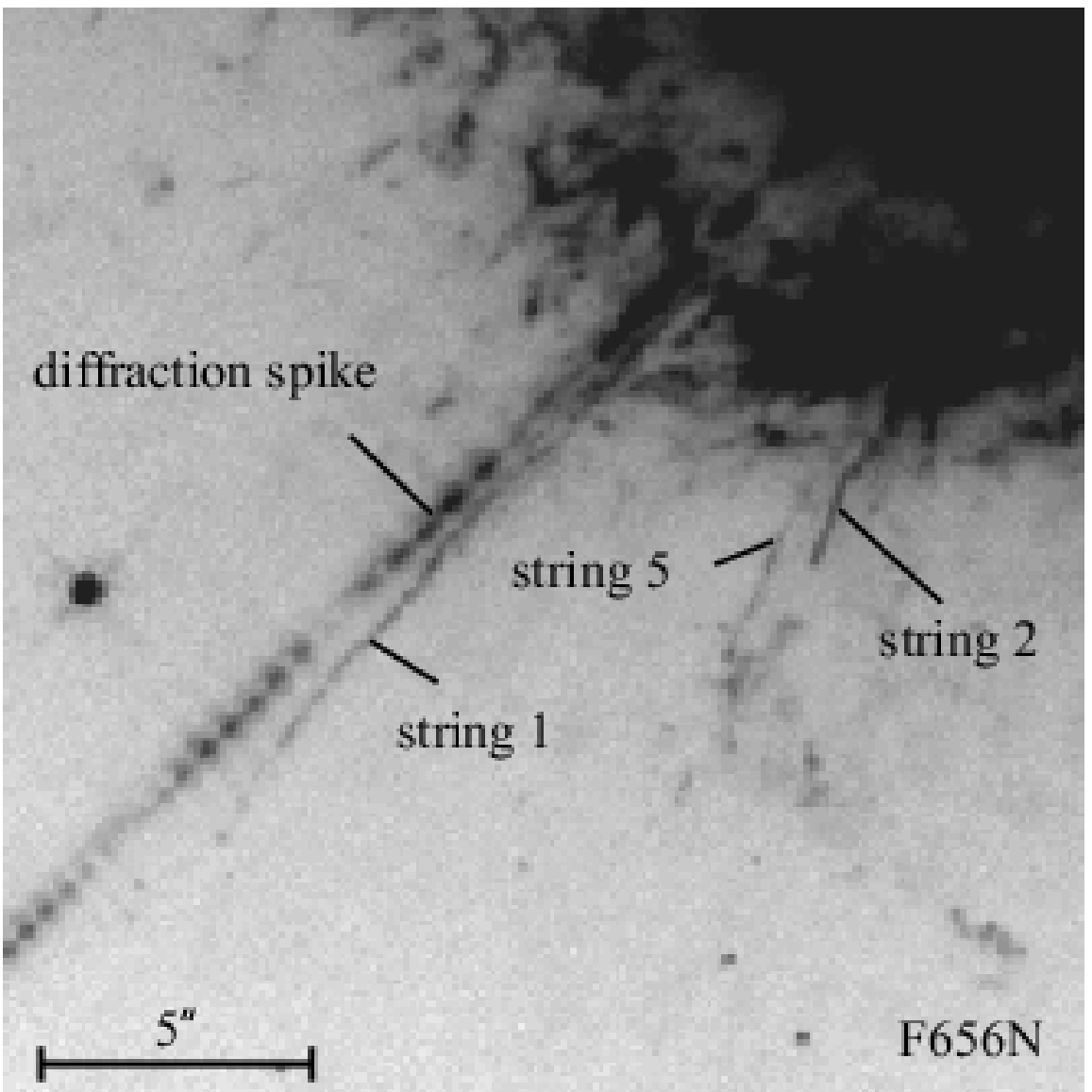}\hspace{0.4cm}\epsfxsize=0.47\textwidth
\epsffile{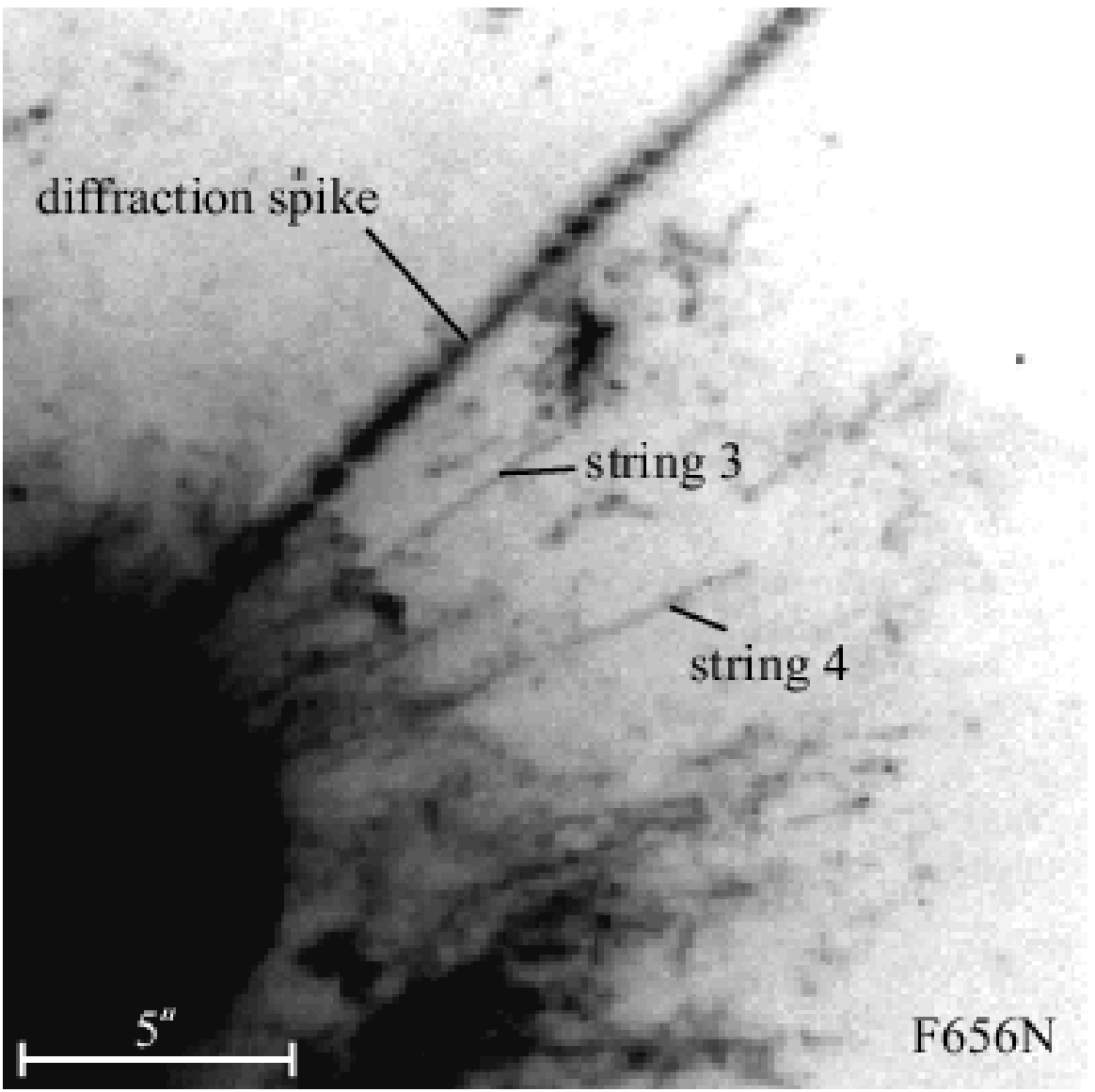}}
\caption{{\it Left:} Enlargement of the F656N HST image
showing the strings 1, 2 and 5 in detail. These three strings are situated
to the south-east of $\eta$ Carinae emerging from the S ridge.
{\it Right:} To the north-east of $\eta$ Carinae,
the strings 3 and 4 are visible.} 
\label{fig:strings}
\end{figure}

Among the structures detected within the outer ejecta are 
the so-called {\it strings}. They are very
straight, highly collimated structures which reach lengths comparable to
the diameter of the entire Homunculus. For a detailed analysis of 
these objects the reader is referred to Weis et al.\ (1999). 
In the nebula around $\eta$ Carinae  
5 strings have been identified by visual inspection of the HST images
(Fig.\,\ref{fig:strings}).
On much smaller scales a large number of string-like objects can be found in 
the outer ejecta. We do not count these smaller structures as strings
since their collimation and parameters are not as extreme as for the strings.

The longest one is string\,1 with an observed length of 15\farcs9 or
0.18\,pc. At the same time string\,1 is only 0\farcs23 wide (0.003\,pc) which
corresponds to a length-to-width ratio of 70, underlining the 
strong collimation of
these objects. String\,1 is bending slightly, and shows no clear `head-like' 
structure but a split within the string of about a length of 
1\farcs5 near the inner (closer to $\eta$ Car) part. 
The other strings are shorter with the smallest
in length being string\,2 with only 4\arcsec\ (Table \ref{tab:stringstab}). 
On the other hand string\,2 is the one with the
highest surface brightness and the only one with a sudden change in its 
(projected) orientation having a kink of about 22\degr. 
Strings\,3 and 4 are very similar to each other in size and appearance, 
they are 7\farcs6
and 9\farcs3 long, respectively, and are extremely straight with nearly no 
variations in surface brightness. String\,5 is the faintest, with its surface
brightness varying the most. 

\begin{figure}
\epsfxsize=0.7\textwidth
\centerline{\epsffile{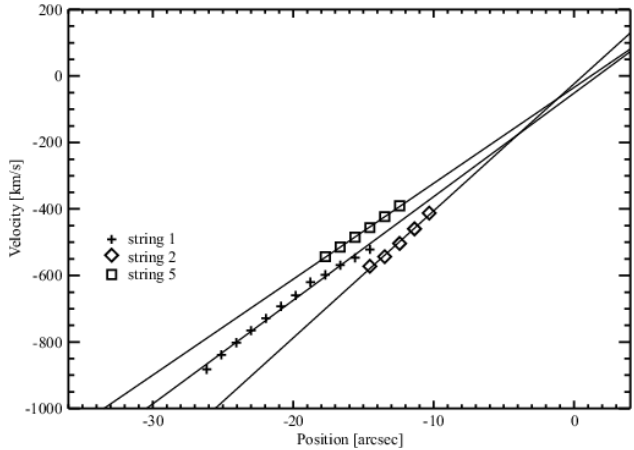}}
\caption{This position-velocity diagram illustrates the constant linear
  increase of the radial velocity of the strings\,1, 2 and 5. Within the errors
  the radial velocity reaches zero if extrapolated back to the position 
of the central star.} 
\label{fig:linear}
\end{figure}

From the long-slit echelle observations we can derive kinematic information
for 3 of the 5 identified strings, namely strings 1, 2 and 5, i.e., the strings
south-east of $\eta$ Carinae (Fig.\,\ref{fig:strings} left).
Analyzing these spectra an amazing kinematic behavior
was discovered. First of all, the strings are moving with relatively high
velocities, the lowest radial velocity measure for string\,1 is 
$-522$\,\kms\ at its innermost end (closest to $\eta$ Car). From there on,
outwards along the string, its radial velocity increases
steadily up to $-995$\,\kms\ at its farest end. The increase of radial
velocity follows a perfectly linear relation, as illustrated in
Fig.\,\ref{fig:linear}. All strings obey this relationship, they only start 
with different velocities. If the radial velocities are extrapolated back
towards the position of the central star the strings would 
reach zero radial velocity there within the errors (see position-velocity 
diagram in Fig.\,\ref{fig:linear}).
The kinematics of all strings can be described with a kind of Hubble-type 
law. The velocity  gradient of each string is slightly
different (see Fig.\,\ref{fig:linear}) and amounts to 
2790\,\kms\,pc$^{-1}$ for string\,1, 3420\,\kms\,pc$^{-1}$ for string\,2 and
2590\,\kms\,pc$^{-1}$ for string\,5. For none of the other strings      
kinematic information were available then. Nevertheless due to the
overall symmetry of the outer ejecta, it was expected that string\,3 and 4 
are redshifted. A first inspection of our recent HST-STIS observations 
of the strings 
(HST-program GO: 8155, PI: Weis) prove that this is the case,  adding a 
further piece of evidence that the strings are also
distributed bipolarly.  

\begin{table}
\caption[]{Properties of the strings}
\begin{center}
\begin{tabular}{cccccc}
\hline
 & observed & observed & length/width\\
\raisebox{1.5ex}[-1.5ex]{string} & length & width & ratio &
\raisebox{1.5ex}[-1.5ex]{$v_{\rm min}$} &
\raisebox{1.5ex}[-1.5ex]{$v_{\rm max}$} \\ 
& [pc] & [10$^{-3}$\,pc] &  & [km\,s$^{-1}$] & [km\,s$^{-1}$] \\ 
\hline 
1    & 0.177 & 3.0 & 70 & $-$522 & $-$995 \\ 
2    & 0.044 & 2.0 & 31 & $-$442 & $-$591 \\ 
3    & 0.095 & 2.0 & 42 & - & - \\ 
4    & 0.103 & 2.0 & 68 & - & - \\ 
5    & 0.058 & 2.0 & 38 & $-$383 & $-$565 \\ \hline
\end {tabular}
\end{center}
\label{tab:stringstab}
\end{table}

The \NH\ ratio for all strings is  3 $\pm$ 0.3 and thus
comparable to the
ratio found in other structures in the nebula around $\eta$ Carinae.

The physical nature of the strings is not well understood yet. They may or may 
not be single physical entities. One may think of a coherent structure, 
similar  to a water jet, for instance. However, one may equally well 
envisage a train of many individual knots or bullets following the same path. 
One also cannot rule out the possibility that they are trails or wakes 
following an object at the strings' far ends, or even projection 
effects of the walls of, for instance, much wider funnels. 
While their high collimation is most likely due to the fact that the strings
move with highly supersonic  velocities, it is harder to explain why 
they follow a 
Hubble type velocity law.  In stellar explosions---rather than winds---a 
linear velocity profile is a good approximation for the larger radii, close
to the head of the explosion (Tscharnuter \& Winkler 1979). One may also
think---if the time scale of the strings' creation 
is short compared to the time scale of their evolution since then---that 
the strings and their expansion velocities can form if this ejection 
happened with a certain distribution of
velocities. Than the fastest part of the string would be the furthest away
from the star, leaving the string as a stratified flow or collection of
bullet with different velocities. If this would be the case 
we needed to explain this distribution of initial velocities
and answer the question why they did  all leave in the same direction?
The creation mechanism and physics of the strings is not solved yet
and it will be of major interest to now analyze our new HST-STIS data
to determine densities and to differentiate between the strings being a flow, 
a chain of bullets, a knot leaving a trail or something 
completely different we have not even thought of yet.

\section{The LBV Nebula around HR Car}

\begin{figure}[t]
\epsfxsize=0.7\textwidth
\centerline{\epsffile{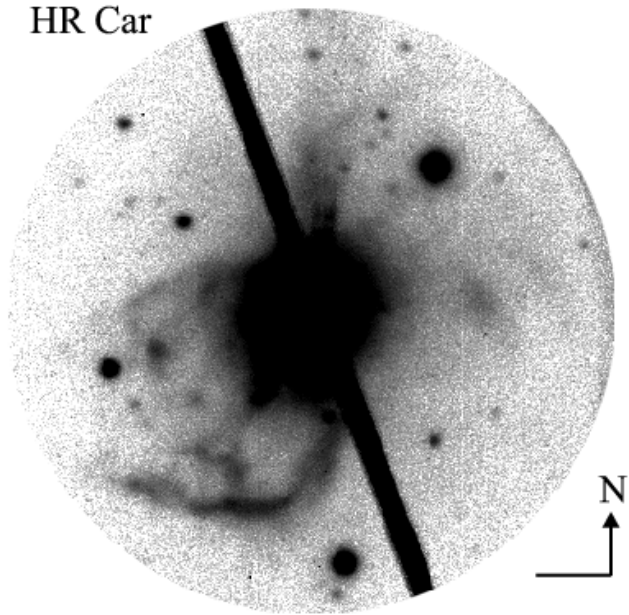}}
\caption{This figure shows an NTT-archive image of the nebula around HR Car 
with the prominent arc in the south-east, which manifests a brighter structure
of the south-eastern lobe.} 
\label{fig:hrcarim}
\end{figure}

While located on the sky closely to $\eta$ Carinae the star HR Carinae  
is about twice the distance from us at
5 $\pm$ 1\,kpc (distance measured by reddening; van Genderen et al.\ 1991).
 Its spectral type varies from B2\,{\sc i} to B9\,{\sc i}.
Strong Balmer, Fe\,{\sc ii} and [Fe\,{\sc ii}] emission lines are observed, 
with the Balmer 
and Fe\,{\sc ii} lines showing P Cygni profiles (Carlson \& Henize 1979; 
Hutsem\'ekers \& Van Drom 1991), i.e.,  a more or less classical spectral
behavior for an LBV star.  HR Car has a luminosity of 
$M_{\rm bol} = -9\fm 0 \pm 0\fm 5$, i.e., at the lower end of the LBV range. 
A circumstellar nebula around HR Car was discovered only in 1991, making 
it one of the newest member of the LBV nebulae school (Hutsem\'ekers \& 
Van Drom 1991).  
The origin and shape of the nebula around HR Car has been discussed by 
Hutsem\'ekers (1994).  A high-resolution image and a spectropolarimetric 
study have been presented by Clampin et al.\ (1995). Based on NTT-archive data
Fig.\,\ref{fig:hrcarim} shows an image of the nebula around HR Car taken
with an H$_{\alpha}$ filter. The central part in this image is occulted 
with a chronographic mask, to avoid the saturation of the bright central star.

The nebula around HR Car was originally classified as a filamentary nebula 
by Nota et al.\ (1995), showing no clear signs of symmetry. 
Nevertheless a careful
inspection of the image in Fig.\,\ref{fig:hrcarim} leads to the suspicion that
the arc in the south-east manifests the brighter part of a lobe 
structure---as do very faint filaments in the north-west. 
The underlying nebula 
may be fainter and reveal a certain symmetry, while the brightest regions 
do not. To decide on the structure of the nebula around HR Car we obtained 
high-resolution echelle spectra similar to
the way we did for $\eta$ Carinae. The detailed description and 
analysis can be found in Weis et al.\ (1997a), here we only summarize the
results. Two long-slit spectra crossing the suspected lobes in their center
(PA=30\degr) offset by 13\arcsec\ north and 
south of the central star
were taken.  In both spectra a Doppler ellipse was detected indicating an 
expansion of $75-150$\,\kms. At both positions a lobe can now be confirmed,
with a diameter of 0.63\,pc for the north-western one and 0.67\,pc for the
south-eastern one. The center of expansion of the Doppler ellipse
of the lobes differ slightly, the north-western lobe is more redshifted, the 
south-eastern lobe is blueshifted most likely due to a tilt of the lobes.
It was therefore proposed by Weis et al.\ (1997a) that HR Car exhibits
a structure morphologically quite similar to the Homunculus nebula 
around $\eta$ Car. 
Missing is the equatorial disk, but note at this location, a very 
bright unresolved central part of the nebula around HR Car 
from where no kinematic data are avaliable yet. This part of the 
nebula might very well hide a disk or at least its remnants.

If we further compare the nebulae around HR Car and $\eta$ Car,
it seems reasonable to propose that the nebula around HR Car 
is an older, evolved version of the nebula around $\eta$ Car.
Each lobe in the HR Car nebula is by a factor of 6 larger than those of $\eta$
Car, and the expansion velocity is by about the same factor lower for HR Car. 
The LBV nebulae around HR Car and $\eta$ Car provide therefore 
an opportunity to probe the evolution of LBV nebulae. 
In such a scenario (Weis et al.\ 1997a) the formation of an 
LBV nebula starts with a nebula that looks more like the Homunculus 
(about 150\,yrs old) and then evolves into a structure similar to
that of HR Car ($4000-9000$\,yrs old).

\section{Bipolar LBV Nebulae---a General Feature?}

\begin{figure}
\epsfxsize=0.9\textwidth
\centerline{\epsffile{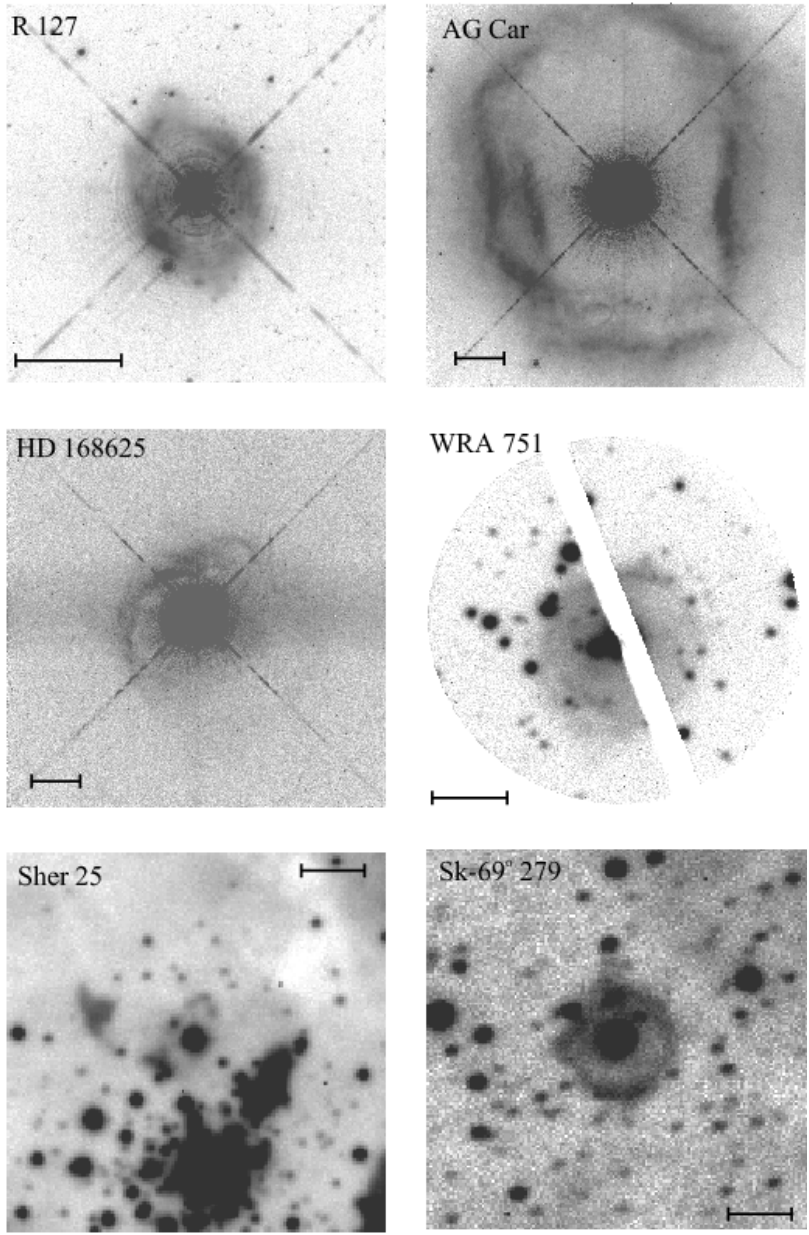}}
\caption{These images show nebula around LBVs and LBV candidates,
which indicate an asymmetric---often bipolar---morphology. 
The scaling bar in each image is 10\arcsec\ in size.}
\label{fig:lbvsbipolar}
\end{figure}

Are $\eta$ Carinae and HR Carinae the only bipolar nebulae among LBVs? 
No---Figure\,\ref{fig:lbvsbipolar} shows a  collection of 
nebulae around LBVs
and LBV candidates. Looking at the individual images and determining the
nebulae's morphology it is obvious that none of the nebulae is really
spherical. All of them show either an elongation or deviation
from spherical symmetry, additional extensions attached to their central body
or even a clearly bipolar structure. If we take R\,127, an LBV in the LMC and
WRA\,751 a galactic LBV we find another similar pair. While their central body
seems spherical both nebulae show almost triangular extensions which we will
call {\it Caps} in the following. For both nebulae it has been shown (Weis
2000a, b) that these Caps are not only 
morphological deformations but 
that they also deviate from the expansion pattern of a sphere. 
In each case the Cap at one side approaches us, while the other Cap is 
receding from us. With respect to the
central star they represent bipolar components. Analyses of the nebula 
around AG Car are manifold (Smith 1991, Nota et al.\ 1992) but still 
no clear 3D structure could be found. The image nevertheless indicates 
that the nebula shows at least a deformation in its center,
forming a `waist-like' structure. In addition it seems that a second shell is
seen in projection underneath or in front of the first shell. For HD 168625 
no good 
kinematic analyses have been published so far---but from the images a 
two-shell structure seems likely (also mentioned by Nota et al.\ 1996), 
probably comparable to the two-shell model for AG Car. 
In Table \ref{table:lbvn} the LBV nebulae known today are listed, as well as
their sizes and expansion velocities. For all the nebulae sizes the diameter is
given or---if asymmetric---at least the diameters along their largest axes. 
Sizes and velocities
separated by a slash indicate that the nebula consists of an inner and an
outer shell and both values are given.

I added two LBV candidates to
the sample in Fig.\,\ref{fig:lbvsbipolar} as well as to 
Table\,\ref{table:lbvn}, the stars Sher\,25 in the galactic H\,{\sc ii} 
region NGC 3603 and Sk$-69\degr$ 279 in the LMC. Both stars 
have not yet been confirmed to be LBVs but their
parameters indicate that they are LBVs indeed.
For Sher\,25 the bipolar structure is already visible from the ground-based
image shown here, but its bipolar nature is even more supported by kinematic 
analyses (Brandner et al.\ 1997a, Brandner et al.\ 1997b). 
Due to the stellar parameters, 
a possible variability and, a for LBVs quite large nebula with an enhanced 
\NH\ ratio, Sher\,25 could be an older LBV (bigger size, smaller expansion 
velocity) like HR Car, putting it towards the end of the evolutionary 
sequence for LBV nebulae. 

The nebula around Sk$-69\degr$ 279 was first reported by Weis et al.\ (1995).
If Sk$-69\degr$ 279 is an LBV, the  nebulae around it would be the 
largest LBV nebula known (4.5\,pc) with the slowest expansion 
velocity of $\approx 14$\,\kms. The highly enhanced \NH\ ratio 
and the star's position in the
HRD make Sk$-69\degr$ 279  another good LBV candidate. This is 
supported by recent analysis of the stars UV spectrum by Smith Neubig \& 
Bruhweiler (1999). They note that the star is characterized by 
parameters similar to that of other LBVs. In addition van Gernderen (2000)
lists the star as an ex/dormant S Dor variable, based on analysis of the star's
variability. Its nebula again 
shows deviations from spherical symmetry especially to the east and  a
very extended filament stretching to the north, which at first look 
is easily mistaken as part of the background H\,{\sc ii} region. Only the 
kinematic data of this object (Weis et al.\ 1997b, Weis \& Duschl 2000)
indicate that this filament is part of the nebula and represents an
outflow of gas. We demonstrated that the LBV S\,119 also 
shows such an outflow (Weis et al.\ 2000a). If outflows in
LBV nebulae are common, this will have important implications for
the determination of kinematic ages of the nebulae and
(determined with help of this method) the true duration of the LBV
phase.

The analysis of the nebulae around $\eta$ Car, HR Car, R\,127 and WRA\,751 
underlines that bipolarity  is present in LBV
nebulae as a quite common feature. It is by far not unique to $\eta$
Carinae, as formerly thought.

\section{Conclusions, Summery and Outlook}

Only recently we begin to understand the evolution of massive stars
in more detail and especially the consequences for the stars'
evolution that result from stellar winds and mass loss. The most massive stars 
encounter the strongest mass loss and therefore their evolution is 
affected the most by the wind. This influence of strong mass loss onto the 
stars' evolution and structure becomes obvious looking at 
stars more massive that about 50\,\Msun. These stars encounter a phase in which
they become unstable, variable and loose even more mass in the LBV phase.
This phase also marks a sudden change in the evolution of the most massive
stars, from redward back to blueward motion in the HRD. LBV nebulae formed in
this phase are the relics and doubtless proofs of the stars' high mass loss.
Studies of various LBV nebulae have been described here---showing the 
difference between as well as strong similarities among these nebulae. 
On the one hand it is obvious that
the nebula around $\eta$ Carinae is special. The extremely 
filamentary outer ejecta, with its very high expansion velocities of up to
2000\,\kms\ is by no doubts outstanding. It has been proven with kinematic
data that beside the bipolar inner Homunculus also the outer ejecta 
is a highly symmetric bipolar nebula. On the other hand this is not as unique
among LBV nebula as other parameters like the expansion 
velocities. An LBV nebula
which very much looks like the nebula around $\eta$ Carinae is that around 
HR Carinae. The bigger size and smaller expansion velocities of the
nebula around HR Car support the idea that HR Car is an older, slowed down 
version of an LBV nebula otherwise similar to that around $\eta$ Car, 
which had time to expand for at least $4000-9000$ years and therefore 
is bigger. 

Quite amazing are the strings in $\eta$ Carinae: Highly collimated, long and
very straight structures expanding with high Mach numbers. Not only 
is it hard to explain how the strings have been accelerated to such high 
velocities, they are also following strictly a linear velocity increase along
the string, they obey a Hubble-type relation. What creates such a structure,
is it a flow, a chain of bullets, a jet, a single knot leaving a trail \dots ? 
A better understanding of these strange objects may be right on the way with
the analysis of our HST-STIS data.  

Looking at the morphology and kinematics of other LBV nebula in our Galaxy as
well as other neighboring galaxies  we found that bipolarity 
in LBV nebulae is indeed a general feature. Different sizes and
expansion velocities of the nebulae can be accounted for by the different ages 
of the nebulae. Differences in their appearance and expansion may easily be
due to interactions with their environments. 
In that case the known LBV nebulae can be described along an evolutionary
sequence. In such a sequence the HR Car nebula, for instance, is an evolved
version of $\eta$ Car's. 

Bipolarity in LBV nebulae is a common feature. 
This poses a new input to the models for stellar
evolution and especially the models for the formation of LBV nebulae. Future
theoretical work has to incooperate the constrains of bipolarity and find mechanism to
conceive this bipolarity. What causes this bipolarity is not known yet. 
Whether stellar rotation, asymmetrical/bipolar winds or a binary
nature could do the job is not clear yet. To decide on that 
further models for the evolution of LBV stars are desperately needed.

\subsection*{Acknowledgements}
The author is very grateful to  Prof.\ Wolfgang J.\ Duschl for 
thoroughly reading and improving the manuscript. Sincere thanks go 
to Dr.\ Dominik J.\ Bomans for several stimulating discussions and 
helpful comments on the subject. This work was supported by the DFG 
through grant Du\,168/8-1. 

\vspace{0.7cm}
\noindent
{\large{\bf References}}
{\small

\bref
Brandner W., Grebel E.K., Chu Y.-H., Weis K., 1997a, ApJ 475, L45

\bref
Brandner W., Chu Y.-H., Eisenhauer F., Grebel E.K., Points S.D, 1997b, ApJ 489,
L153

\bref
Carlson E.D., Henize K.G., 1979, Vistas in Astron. 23, 213

\bref
Chiosi C., Maeder A., 1986, ARA\&A 24, 329

\bref
Clampin M., Schulte-Ladbeck R.E., Nota A.\ et al., 1995, AJ 110, 251

\bref
Currie D.G., Dowling D.M., Shaya E.J., et al. 1996, AJ 112, 1115

\bref
Currie D.G., Dowling D.M., 1999, in {\it Eta Carinae at the Millennium},
ASP Conf. Ser.\ 179, eds.: Morse J.A., Humphreys R.M., Damineli A.,p. 72

\bref
Damineli A., 1996, ApJ 460, L49

\bref
Damineli A., Conti P.S., Lopes D.F., 1997, New Astronomy 2, 107

\bref
Davidson K., Humphreys R.M., 1997, ARA\&A 35, 1

\bref
de Jager C., Nieuwenhuijzen H., van der Hucht K.A., 1988, A\&AS 72, 259

\bref
Duschl W.J., Hofmann K.-H., Rigaut F., Weigelt G., 1995, RevMexAA SdC 2, 17 

\bref
Figer D.F., Morris M., Geballe T.R., Rich R.M., Serabyn E., McLean I.S.,
Puetter R.C., Yahil A., 1999, ApJ 525, 759

\bref
Garc{\'\i}a-Segura G., Mac Low M.-M., Langer N., 1996, A\&A 305, 229

\bref
Gaviola E., 1946, Revista Astronomica 18, 25

\bref
Gaviola E., 1950, ApJ 111, 408

\bref
Gratton L., 1963, in {\it Star Evolution}, eds.: Gratton L., New York 
Academic Press, p. 297

\bref
Humphreys R.M., 1999, in  Proc. IAU Collq. No. 169, {\it Variable and 
Non-spherical Stellar Winds in Luminous 
Hot Stars}, eds.: Wolf B., Stahl O., Fullerton A.W., Lecture Notes in 
Physics, Springer, p. 243

\bref
Humphreys R.M., Davidson K., 1979, ApJ 232, 409

\bref
Humphreys R.M., Davidson K., 1994, PASP 106, 1025

\bref
Humphreys R.M., Davidson K., Smith N., 1999, PASP, 111, 1124 

\bref
Hutsem\'ekers D., 1994, A\&A 281, L81

\bref
Hutsem\'ekers D., Van Drom E., 1991, A\&A 248, 141

\bref
Kudritzki R.P., 1999, in  Proc. IAU Collq. No. 169, {\it Variable and 
Non-spherical Stellar Winds in Luminous 
Hot Stars}, eds.: Wolf B., Stahl O., Fullerton A.W., Lecture Notes in 
Physics, Springer, p. 405

\bref
Kudritzki R.P., Lennon D.J., Haser S.M., Puls J., Pauldrach A.W.A., Venn K.,
Voels S.A., 1996, in {\it Science with the Hubble Space Telescope II},
eds.: Benvenuti P. et al., p. 285

\bref
Langer N., Hamann W.-R., Lennon M., Najarro F., Pauldrach A.W.A., Puls J.,
1994, A\&A 290, 819

\bref
Maeder A., Conti P.S., 1994, ARA\&A 32, 227

\bref
Maeder A., Meynet G., 2000, A\&A 361, 159

\bref
Massey P., Johnson J., 1993, AJ 105, 980

\bref
Meaburn J., Wolstencroft R.D., Walsh J.R., 1987, A\&A 181, 333

\bref
Meaburn J., Walsh J.R., Wolstencroft R.D., 1993a, A\&A 268, 283

\bref
Meaburn J., Gehring G., Walsh J.R. et al. 1993b, A\&A 276, L21

\bref 
Meaburn J., L{´o}pez J.A., Barlow M.J., Drew J.E., 1996a, MNRAS 283, L69

\bref
Meaburn J., Boumis P., Walsh J.R. et al., 1996b, MNRAS 282, 1313

\bref
Morse J.A., Davidson K., Bally J., et al., 1998, AJ 116, 2443

\bref 
Nota A., Leithere C., Clampin M., Greenfield P., Golimowski D.A., 1992, ApJ
398, 621

\bref
Nota A., Livio M., Clampin M., Schulte-Ladbeck R., 1995, ApJ 448, 788

\bref
Nota A., Pasquali A., Clampin M., Pollacco D., Scuderi S., Livio M., 1996, 
ApJ 473, 946

\bref
Nota A., Smith L.J., Pasquali A., Clampin M., Stroud M., 1997, ApJ 486, 338

\bref
Ringuelet A.E., 1958, Zeitschrift f\"ur Astrophysik, Bd. 46, p.276

\bref
Schaerer D., de Koter A., Schmutz W., Maeder A., 1996a, A\&A 310, 837

\bref
Schaerer D., de Koter A., Schmutz W., Maeder A., 1996b, A\&A 312, 475

\bref
Schaller G., Schaerer D., Meynet G., Maeder A., 1992, A\&AS 96, 269

\bref
Smith L.J., 1991 in IAU Symp. 143, {\it Wolf-Rayet Stars and Interrelations 
with Other Massive Stars in Galaxies}, eds.: van der Hucht K.A., Hidayat B.,
Kluwer, Dordrecht, Holland, p. 385

\bref
Smith L.J., Crowther P.A., Prinja R.K., A\&A 281, 833

\bref
Smith N., Gehrz R.D., 1998, AJ 116, 823

\bref
Smith Neubig M.M., Bruhweiler F.C., 1999, AJ 117, 2856

\bref
Stahl O., Damineli A., 1998, in {\it Cyclical Variability in stellar winds},
proceedings of the ESO workshop, eds.: Kaper L., Fullerton A.W., 
Springer, p. 112

\bref
Thackeray A.D., 1949, Observatory 69, 31

\bref
Thackeray A.D., 1950, MNRAS 110, 524

\bref
Thackeray A.D., 1961, Obs. 81, 99

\bref
Tscharnuter W.M., Winkler K.-H., 1979, Comp.Phys.Comm.\ 18, 171

\bref
van den Bos, 1938, Union Observatory Circular, No. 100, p.522  

\bref
van Genderen A.M., 2000, A\&AS, in press

\bref
van Genderen A.M., Robijn F.H.A., van Esch B.P.M., Lamers H.J.G.L.M., 1991, 
A\&A 246, 407

\bref
Viotti R., 1995, RevMexAA SdC\,2, 10

\bref
Walborn N.R., 1976, ApJ 204, L17

\bref
Walborn N.R., Evans I.N., Fitzpatrick E.L., Phillips M.M., 1991,
in IAU Symp. 143, {\it Wolf-Rayet Stars and Interrelations 
with Other Massive Stars in Galaxies}, Hrsg.: van der Hucht K.A., Hidayat B.,
Kluwer, Dordrecht, Holland, p. 505

\bref
Walborn N.R., 1995, RevMexAA SdC\,2, 51

\bref
Walborn N.R., Blanco B.M., Thackeray A.D., 1978, ApJ 219, 498 

\bref
Walborn N.R., Blanco B.M., 1988, PASP 100, 797

\bref
Weis K., 2000a, A\&A 357, 938 

\bref
Weis K., 2000b, A\&A submitted 

\bref
Weis K. Bomans D.J., Chu Y.-H., Joner M.D., Smith R.C., 1995, RevMexAA (SC) 3,
237

\bref
Weis K., Duschl W.J., Bomans D.J., Chu Y.-H., Joner M.D., 1997a, A\&A 320, 568

\bref
Weis K., Chu Y.-H.,  Duschl W.J., Bomans D.J., 1997b, A\&A 325, 1157

\bref
Weis K., Duschl W.J., Chu Y.-H., 1999, A\&A 349, 467 

\bref
Weis K., Duschl W.J., 2000, A\&A submitted

\bref
Weis K., Duschl W.J., Bomans D.J., 2000a, A\&A submitted

\bref
Weis K., Duschl W.J., Bomans D.J., 2000b, A\&A in press and
astro-ph/0012426
}
\vfill

\end{document}